\documentstyle[12pt]{article}
\input epsfig

\begin{document}
\baselineskip=20.5pt
\def\beqra{\begin{eqnarray}} \def\eeqra{\end{eqnarray}}
\def\beqast{\begin{eqnarray*}} \def\eeqast{\end{eqnarray*}}
\def\beq{\begin{equation}}      \def\eeq{\end{equation}}
\def\be{\begin{enumerate}}   \def\ee{\end{enumerate}}

\def\fnote#1#2{\begingroup\def\thefootnote{#1}\footnote{#2}\addtocounter
{footnote}{-1}\endgroup}

\def\gam{\gamma}
\def\Gam{\Gamma}
\def\la{\lambda}
\def\eps{\epsilon}
\def\La{\Lambda}
\def\si{\sigma}
\def\Si{\Sigma}
\def\al{\alpha}
\def\Tha{\Theta}
\def\tha{\theta}
\def\vphi{\varphi}
\def\del{\delta}
\def\Del{\Delta}
\def\ab{\alpha\beta}
\def\om{\omega}
\def\Om{\Omega}
\def\mn{\mu\nu}
\def\mun{^{\mu}{}_{\nu}}
\def\kap{\kappa}
\def\rsi{\rho\sigma}
\def\beal{\beta\alpha}
\def\til{\tilde}
\def\rta{\rightarrow}
\def\eqv{\equiv}
\def\nab{\nabla}
\def\pa{\partial}
\def\sit{\tilde\sigma}
\def\ul{\underline}
\def\indt{\parindent2.5em}
\def\nd{\noindent}
\def\rsi{\rho\sigma}
\def\beal{\beta\alpha}
\def\caa{{\cal A}}
\def\cb{{\cal B}}
\def\cac{{\cal C}}
\def\cd{{\cal D}}
\def\ce{{\cal E}}
\def\cf{{\cal F}}
\def\cg{{\cal G}}
\def\cah{{\cal H}}
\def\ci{{\cal I}}
\def\cj{{\cal{J}}}
\def\ck{{\cal K}}
\def\cl{{\cal L}}
\def\cm{{\cal M}}
\def\cn{{\cal N}}
\def\cO{{\cal O}}
\def\cp{{\cal P}}
\def\car{{\cal R}}
\def\cs{{\cal S}}
\def\ct{{\cal{T}}}
\def\cu{{\cal{U}}}
\def\cv{{\cal{V}}}
\def\cw{{\cal{W}}}
\def\cx{{\cal{X}}}
\def\cy{{\cal{Y}}}
\def\cz{{\cal{Z}}}
\def\asymptotic{{_{\stackrel{\displaystyle\longrightarrow}
{x\rightarrow\pm\infty}}\,\, }} 
\def\asymptext{\raisebox{.6ex}{${_{\stackrel{\displaystyle\longrightarrow}
{x\rightarrow\pm\infty}}\,\, }$}} 
\def\asymptoticp{{_{\stackrel{\displaystyle\longrightarrow}
{x\rightarrow +\infty}}\,\, }} 
\def\asymptoticm{{_{\stackrel{\displaystyle\longrightarrow}
{x\rightarrow -\infty}}\,\, }} 

\def\raisenot{\raise .5mm\hbox{/}}
\def\nota{\ \hbox{{$a$}\kern-.49em\hbox{/}}}
\def\notA{\hbox{{$A$}\kern-.54em\hbox{\raisenot}}}
\def\notb{\ \hbox{{$b$}\kern-.47em\hbox{/}}}
\def\notB{\ \hbox{{$B$}\kern-.60em\hbox{\raisenot}}}
\def\notc{\ \hbox{{$c$}\kern-.45em\hbox{/}}}
\def\notd{\ \hbox{{$d$}\kern-.53em\hbox{/}}}
\def\notbd{\ \hbox{{$D$}\kern-.61em\hbox{\raisenot}}} 
\def\note{\ \hbox{{$e$}\kern-.47em\hbox{/}}}
\def\notk{\ \hbox{{$k$}\kern-.51em\hbox{/}}}
\def\notp{\ \hbox{{$p$}\kern-.43em\hbox{/}}}
\def\notq{\ \hbox{{$q$}\kern-.47em\hbox{/}}}
\def\notW{\ \hbox{{$W$}\kern-.75em\hbox{\raisenot}}}
\def\notz{\ \hbox{{$Z$}\kern-.61em\hbox{\raisenot}}}
\def\notpa{\hbox{{$\partial$}\kern-.54em\hbox{\raisenot}}}

\def\fo{\hbox{{1}\kern-.25em\hbox{l}}}  
\def\rf#1{$^{#1}$}
\def\bx{\Box}
\def\tr{{\rm Tr}}
\def\rmtr{{\rm tr}}
\def\dgg{\dagger}
\def\lag{\langle}
\def\rag{\rangle}
\def\bmid{\big|}
\def\vlap{\overrightarrow{\La p}} 
\def\lrta{\longrightarrow} \def\lrar{\raisebox{.8ex}{$\longrightarrow$}}
\def\ON{{\cal O}(N)}
\def\UN{{\cal U}(N)}
\def\bdPh{\mbox{\boldmath{$\dot{\!\Phi}$}}}
\def\bPh{\mbox{\boldmath{$\Phi$}}}
\def\bPhs{\bPh^2}
\def\sef{S_{eff}[\sigma,\pi]}
\def\sigx{\sigma(x)}
\def\pix{\pi(x)}
\def\bph{\mbox{\boldmath{$\phi$}}}
\def\bphs{\bph^2}
\def\ex{\BM{x}}
\def\exs{\ex^2}
\def\xdot{\dot{\!\ex}}
\def\y{\BM{y}}
\def\ys{\y^2}
\def\ydot{\dot{\!\y}}
\def\pat{\pa_t}
\def\pax{\pa_x}

\renewcommand{\thesection}{\arabic{section}}
\renewcommand{\theequation}{\thesection.\arabic{equation}}


\vspace*{.2in}
\begin{center}
  \Large{\sc The Dirac Operator in a Fermion Bag Background in $1+1$ 
Dimensions and Generalized Supersymmetric Quantum Mechanics}\\
\normalsize
\vspace{15pt}
\begin{center}
{\bf Joshua Feinberg
\fnote{*}{{\it e-mail address: joshua@physics.technion.ac.il}}}
\end{center}
\vskip 2mm
\begin{center}
{Physics Department,}\\
{University of Haifa at Oranim, Tivon 36006, Israel\fnote{**}{{\it permanent address}}}\\
{and}\\
{Physics Department,}\\
{Technion, Israel Institute of Technology, Haifa 32000, Israel}\\
\vskip 2mm
\end{center}
\vspace{.3cm}
\end{center}

\begin{minipage}{5.3in}
{\abstract~~~~~
We show that the spectral theory of the Dirac operator 
$D = i\notpa-\si (x)-i\pi (x) \gam_5$ in a static background 
$(\sigx, \pix)$ in $1+1$ space-time dimensions, is underlined by 
a certain generalization of supersymmetric quantum mechanics, 
and explore its consequences.} 
\end{minipage}

\vspace{10pt}

\vfill
\pagebreak

\setcounter{page}{1}

\section{Introduction: Bags and Resolvents}
A central concept in particle physics states that
fundamental particles acquire their masses through interactions with vacuum
condensates. Thus, a massive particle may carve out around itself a spherical
region \cite{sphericalbag} or a shell \cite{shellbag} in which the
condensate is suppressed, thus reducing the effective mass of the particle
at the expense of volume and gradient energy associated with the
condensate. This picture has interesting phenomenological consequences
\cite{sphericalbag, mackenzie}.

This phenomenon may be studied non-perturbatively in model field theories 
in $1+1$ space-time dimensions such as the Gross-Neveu (GN) model \cite{gn}
and the multi-flavor Nambu-Jona-Lasinio (NJL) \cite{njl} model,
in the large $N$ limit.

Explicit calculations of fermion bag profiles in the GN and NJL models were 
given originally in \cite{ccgz}, \cite{dhn} and in \cite{shei}.

Following these works, fermion bags in the GN and NJL models were 
discussed in the literature several other times \cite{others}, \cite{josh1},
\cite{fz}, \cite{massivegn}. For a recent review on these and related 
matters, see \cite{thies}.

Very recently, static chiral fermion bag solitons \cite{jaffe} in a $1+1$ 
dimensional model, as well as non-chiral (real scalar) fermion bag solitons 
\cite{bashinsky}, were discussed, in which the scalar field that couples to 
the fermions was dynamical already at the classical level.

Mathematical considerations similar to those involved in studying
fermion bags, appear also in other branches of theoretical physics,
such as the theory of inhomogeneous superconductors \cite{stone},
and the results of this paper may be applicable there as well.

Studying the physics of fermion bags necessarily involves knowledge of the
resolvent of the Dirac operator in the background of the bag. As an example, 
let us consider the $1+1$ dimensional NJL model
(which contains the GN model as a special case).

One version of writing the action of the $1+1$ dimensional NJL model is 
\beq
S=\int d^2x\,\left\{\sum_{a=1}^N\, \bar\psi_a\,\left[i\notpa-(\si+i\pi\gam_5)
\right]\,\psi_a 
-{1\over 2g^2}\,(\si^2+\pi^2)\right\}\,,
\label{auxiliary}
\eeq
where the $\psi_a\,(a=1,\ldots,N)$ are $N$ flavors of massless Dirac 
fermions, with Yukawa couplings to the scalar and pseudoscalar auxiliary 
fields $\si(x), \pi(x)$.

The partition function associated with (\ref{auxiliary}) 
is
\beq
\cz=\int\,\cd\si\,\cd\pi\,\cd\bar\psi\,\cd\psi \,\exp i\, 
\int\,d^2x\left\{\bar\psi
\left[i\notpa-\left(\si+i\pi\gam_5\right)\right]\psi-{1\over 
2g^2}\,\left(\si^2+\pi^2\right)\right\}
\label{partition}
\eeq
Integrating over the grassmannian variables leads to 
$\cz=\int\,\cd\si\,\cd\pi\,\exp \{iS_{eff}[\si,\pi]\}$
where the bare effective action is
\beq
S_{eff}[\si,\pi] =-{1\over 2g^2}\int\, d^2x 
\,\left(\si^2+\pi^2\right)-iN\, 
\tr\log\left[i\notpa-\left(\si+i\pi\gam_5\right)\right]
\label{effective}
\eeq
and the trace is taken over both functional and Dirac indices. 

This theory has been studied in the limit 
$N\rightarrow\infty$ with $Ng^2$ held fixed\cite{gn}. In this limit, 
(\ref{partition}) is governed by saddle points of (\ref{effective}) 
and the small fluctuations around them. The most general saddle point
condition reads

\beqra
{\del S_{\em eff}\over \del \si\left(x,t\right)}  &=&
-{\si\left(x,t\right)\over g^2} + iN ~{\rm tr} \left[~~~~~ \langle x,t | 
{1\over i\notpa
-\left(\si + i\pi\gam_5\right)} | x,t \rangle \right]= 0
\nonumber\\{}\nonumber\\
{\del S_{\em eff}\over \del \pi\left(x,t\right)}  &=&
-{\pi\left(x,t\right)\over g^2} - ~N~ {\rm tr} \left[~\gam_5~\langle x,t 
| {1\over i\notpa
-\left(\si + i\pi\gam_5\right)} | x,t \rangle~\right] = 0\,.
\label{saddle}
\eeqra
Fermion bags are the space-time dependent solutions
$(\si\left(x,t\right), \pi\left(x,t\right))$ of (\ref{saddle}),
subjected to appropriate boundary conditions at spatial infinity, 
and on which  $S_{eff}/N$ is finite.

Thus, studying fermion bags necessarily involves the
resolvent of the Dirac operator in the background of the bag.

In this paper we discuss some mathematical aspects of the much 
simpler problem of {\em static} fermion bags, namely, the static solutions 
$(\sigx, \pix)$ of (\ref{saddle}).

For the usual physical reasons, we set boundary conditions on our static 
fields such that $\sigx$ and $\pix$ start from a point on the 
vacuum manifold $\si^2 + \pi^2 = m^2$ (with constant $\si,\pi$ of course, 
and where $m$ is the dynamical mass \cite{gn} ) at $x=-\infty$, wander 
around in the
$\si-\pi$ plane, and then relax back to another point on the vacuum manifold
at $x=+\infty$. Thus, we must have the asymptotic behavior 
\beqra
&&\si\asymptotic m{\rm cos}\theta_{\pm}\quad\quad ,\quad\quad 
\si'\asymptotic 0 
\nonumber\\{}\nonumber\\
&&\pi\asymptotic m{\rm sin}\theta_{\pm}\quad\quad , 
\quad\quad \pi'\asymptotic 0
\label{boundaryconditions}
\eeqra
where $\theta_{\pm}$ are the asymptotic chiral alignment angles. Only the 
difference $\theta_+ - \theta_-$ is meaningful, of course, and henceforth 
we use the axial $U(1)$ symmetry of (\ref{auxiliary}) to set 
$\theta_- = 0$, such that $\si (-\infty)=m$ and $\pi (-\infty)=0$. 
We also omit the subscript from $\theta_+$
and denote it simply by $\theta$ from now on. As typical of solitonic
configurations, we expect, that $\sigx$ and $\pix$ tend to their asymptotic 
boundary values (\ref{boundaryconditions}) on the vacuum manifold at an 
exponential rate which is determined, essentially, by the mass gap $m$ of 
the model.

Thus, in order to study static fermion bags, we need to invert the Dirac 
operator 
\beq
D\equiv i\notpa-(\sigx+i\pix\gam_5)
\label{dirac}
\eeq
in a given background of static field configurations $\sigx$ and $\pix$,
subjected to the boundary conditions (\ref{boundaryconditions}). 
In particular, we have to find the diagonal resolvent of (\ref{dirac}) in 
that background. We stress that inverting (\ref{dirac}) has nothing to do 
with the large $N$ approximation, and consequently our results 
are valid for any value of $N$.

The rest of the paper is organized as follows: In Section 2 we show that 
the Dirac equation $\left(i\notpa-(\sigx+i\pix\gam_5)\right)\psi=0$
in a given static $\sigx + i\gam_5\pix$ background,
is equivalent to a pair of two isospectral Sturm-Liouville equations 
in one dimension, which generalize the well known one-dimensional 
supersymmetric quantum mechanics.
We use this generalized supersymmetry to express all 
four entries of the space-diagonal Dirac resolvent 
(i.e., the resolvent evaluated at coincident spatial coordinates) in terms 
of a single function. In Section 3, we use the results of Section 2 to derive
simple expressions for various bilinear fermion condensates in the 
given static $\sigx + i\gam_5\pix$ background. In particular, 
we prove that each frequency mode of the spatial current 
$\langle \bar\psi (x) \gam^1 \psi (x)\rangle$ 
vanishes identically in the static background. 
\pagebreak

\section{Resolvent of the Dirac Operator With Static Background Fields }
\setcounter{equation}{0}

As was explained in the introduction, we need to invert the Dirac operator 
(\ref{dirac}), $D\equiv i\notpa-(\sigx+i\pix\gam_5)$,
in a given background of static field configurations $\sigx$ and $\pix$,
subjected to the boundary conditions (\ref{boundaryconditions}).

In this paper we use the Majorana
representation  
\beq\label{majorana}
\gam^0=\si_2\;,\; \gam^1=i\si_3\quad {\rm and} \quad 
\gam^5=-\gam^0\gam^1=\si_1
\eeq
for $\gam$ matrices. In this representation  (\ref{dirac}) becomes
\beq
D =\left(\begin{array}{cc} -\pa_x - \si & -i\omega - i\pi \\{}&{}\\ 
i\omega- i\pi & \pa_x - \si\end{array}\right)
=\left(\begin{array}{cc} -Q & -i\omega - i\pi \\{}&{}\\ 
i\omega- i\pi & -Q^\dgg\end{array}\right)\,,
\label{dirac1}
\eeq
where we introduced the pair of adjoint operators 
\beq\label{qqdagger}
Q = \sigx + \pa_x\,,\quad\quad Q^\dgg = \sigx - \pa_x\,.
\eeq
(To obtain (\ref{dirac1}), we have naturally transformed 
$i\notpa-(\sigx+i\pix\gam_5)$ to the $\om$ plane, since the 
background fields $\sigx, \pix$ are static.) 

Inverting (\ref{dirac1}) is achieved by solving  
\beq
\left(\begin{array}{cc} -Q & -i\omega - i\pix \\{}&{}\\ 
i\omega- i\pix & -Q^\dgg\end{array}\right)\cdot 
\left(\begin{array}{cc} a(x,y) &  b(x,y) \\{}&{}\\ c(x,y) & 
d(x,y)\end{array}\right)\,=\,-i{\bf 1}\del(x-y)
\label{greens}
\eeq
for the Green's function of (\ref{dirac1}) in a given background 
$\sigx,\pix$.
By dimensional analysis, we see that the quantities $a,b,c$ and 
$d$ are dimensionless.

\subsection{Generalized ``Supersymmetry'' in a Chiral Bag Background}
We now show that the spectral theory of the Dirac operator (\ref{dirac1})
is underlined by a certain generalized one dimensional supersymmetric 
quantum mechanics. This generalized supersymmetry
is very helpful in simplifying various calculations involving the Dirac 
operator and its resolvent.

The diagonal elements $a(x,y), ~d(x,y)$ in (\ref{greens}) may be 
expressed
in term of the off-diagonal elements as
\beq
a(x,y)={-i\over \omega-\pix}Q^\dgg c(x,y)\,,\quad\quad
d(x,y)={i\over \omega+\pix} Q b(x,y)
\label{ad}
\eeq
which in turn satisfy the second order partial differential equations
\beqra
&&\left[Q^\dgg {1\over \om+\pix}Q -(\om-\pix )\right]b(x,y)\,=
\nonumber\\&&{}\nonumber\\
&&-\pa_x\left[{\pa_x b(x,y)\over 
\omega+\pix}\right]+\left[\sigx^2+\pix^2-\si'(x)-\omega^2+{\sigx\pi'(x)
\over 
\omega+\pix}\right]{b(x,y)\over \omega+\pix}\,=\,~~\del(x-y)
\nonumber\\&&{}\nonumber\\
&&\left[Q {1\over \om-\pix}Q^\dgg -(\om+\pix )\right]c(x,y)\,=
\nonumber\\&&{}\nonumber\\
&&-\pa_x\left[{\pa_x c(x,y)\over 
\omega-\pix}\right]+\left[\sigx^2+\pix^2+\si'(x)-\omega^2+{\sigx\pi'(x)
\over 
\omega-\pix}\right]{c(x,y)\over 
\omega-\pix}\,=\,-\del(x-y)\,.\nonumber\\&&{}
\label{bc}
\eeqra

Thus, $b(x,y)$ and $-c(x,y)$ are simply the Green's functions of the 
corresponding second order Sturm-Liouville operators\footnote{Note that 
$\om$ plays here a dual role: in addition to its role as the spectral 
parameter (the $\om^2$ terms in (\ref{bcops})), it also appears as a
parameter in the definition of these operators-hence the explicit $\om$ 
dependence in our notations for these operators in (\ref{bcops}). For
this reason, it is possible to completely factorize the operators $L_b$ 
and $L_c$ by additional obvious $\om$-dependent similarity transformations 
on $Q$ and $Q^\dgg$. However, these similarity transformations are singular 
at points where $\pix = \pm\om$ and are thus ill defined, and we will avoid 
them.}
\beqra\label{bcops}
&& L_b(\om) b(x) = -\pa_x\left[{\pa_x b(x)\over 
\omega+\pix}\right]+\left[\sigx^2+\pix^2-\si'(x)-\omega^2+{\sigx\pi'(x)
\over 
\omega+\pix}\right]{b(x)\over \omega+\pix}
\nonumber\\&&{}\nonumber\\
&& L_c(\om) c(x) = -\pa_x\left[{\pa_x c(x)\over 
\omega-\pix}\right]+\left[\sigx^2+\pix^2+\si'(x)-\omega^2+{\sigx\pi'(x)
\over 
\omega-\pix}\right]{c(x)\over 
\omega-\pix}\nonumber\\&&{}
\eeqra
in (\ref{bc}), namely, 
\beqra
b(x,y)&=&{\theta\left(x-y\right)b_2(x)b_1(y)+\theta\left(y-x\right)b_2(y)
b_1(x) \over W_b}
\nonumber\\{}\nonumber\\
c(x,y)&=&-{\theta\left(x-y\right)c_2(x)c_1(y)+\theta\left(y-x\right)c_2(y
)c_1(x)\over W_c}\,.
\label{bcexpression}
\eeqra
Here $\{b_1(x), b_2(x)\}$ and $\{c_1(x), c_2(x)\}$ are pairs of independent 
fundamental solutions of the two equations $L_b b(x)=0$ and $L_c c(x)=0$, 
subjected to the 
boundary conditions 
\beq\label{planewaves}
b_1(x)\,,c_1(x) \asymptoticm A_{b,c}^{(1)}(k)e^{-ikx} \quad\quad ,\quad\quad 
b_2(x)\,,c_2(x)\asymptoticp A_{b,c}(k)^{(2)}e^{ikx}
\eeq
with some possibly $k$ dependent coefficients $A_{b,c}^{(1)}(k), 
A_{b,c}^{(2)}(k)$ and with\footnote{We see that if ${\rm Im}k>0$, $b_1$ 
and $c_1$ decay exponentially to the left, and $b_2$ and $c_2$ decay to 
the right. Thus, if ${\rm Im}k>0$, both  $b(x,y)$ and $c(x,y)$ decay as 
$|x-y|$ tends to infinity.} 
\beq\label{kmom}
k=\sqrt{\om^2 -m^2}\,,\quad {\rm Im}k\geq 0\,.
\eeq
The purpose of introducing the (yet unspecified) 
coefficients $A_{b,c}^{(1)}(k), A_{b,c}^{(2)}(k)$ 
will become clear following Eqs. (\ref{btoc}) and (\ref{ctob}). 
The boundary conditions (\ref{planewaves}) are consistent, of course, with 
the asymptotic behavior (\ref{boundaryconditions}) of $\si$ and $\pi$ due to 
which both $L_b$ and $L_c$ tend to a free particle hamiltonian 
$[-\pa_x^2 + m^2-\om^2]$ as $x\rightarrow\pm\infty$.

The wronskians of these pairs of solutions are 
\beqra\label{wronskian}
&& W_b(k) ={b_2(x)b_1^{'}(x)-b_1(x)b_2^{'}(x)\over \omega+\pix}
\nonumber\\&&{}\nonumber\\
&& W_c(k) = {c_2(x)c_1^{'}(x)-c_1(x)c_2^{'}(x)\over \omega-\pix}
\nonumber\\&&{}
\eeqra
As is well known, $W_b(k)$ and $W_c (k)$ are independent of $x$.

Note in passing that the canonical asymptotic behavior assumed in 
the scattering theory of the operators $L_b$ and $L_c$ corresponds to setting 
$A_{b,c}^{(1)} = A_{b,c}^{(2)} = 1$ in (\ref{planewaves}). Thus, 
the wronskians in (\ref{wronskian}) are {\em not} the canonical wronskians 
used in scattering theory. As is well known in the literature \cite{faddeev},
the {\em canonical} wronskians are proportional 
(with a $k$ independent coefficient) to $k/t(k)$, where $t(k)$ is 
the transmission amplitude of the corresponding 
operator $L_b$ or $L_c$. Thus, on top of the well-known features of $t(k)$, 
such as the fact that $t(k)$ has simple poles on the positive imaginary 
$k$-axis (corresponding to bound states), the wronskians in 
(\ref{wronskian}) will have additional spurious 
$k$-dependence coming from the amplitudes 
$A_{b,c}^{(1)}(k), A_{b,c}^{(2)}(k)$ in (\ref{planewaves}).

Substituting the expressions (\ref{bcexpression}) for the off-diagonal 
entries $b(x,y)$ and $c(x,y)$ into (\ref{ad}), we obtain the appropriate 
expressions for the diagonal entries $a(x,y)$ and $d(x,y)$. We do not bother
to write these expressions here. It is useful however to note, that despite 
the $\pax$'s in the $Q$ operators in (\ref{ad}), that act on the 
step functions in (\ref{bcexpression}), neither $a(x,y)$ nor $d(x,y)$ contain 
pieces proportional to $\del(x-y)$\,. Such pieces cancel one another due 
to the symmetry of (\ref{bcexpression}) under $x\leftrightarrow y$.

We will now prove that the spectra of the operators $L_b$ and $L_c$ 
are essentially the same. Our proof is based on the fact that we can 
factorize the eigenvalue equations $L_b b(x)=0$ and $L_c c(x)=0$ as
\beqra\label{bcfactor}
&& {1\over \om-\pix}\,Q^\dgg\, {1\over \om +\pix}\, Q\, b = b
\nonumber\\&&{}\nonumber\\
&& {1\over \om+\pix}\,Q\, {1\over \om -\pix}\, Q^\dgg\, c = c\,,
\nonumber\\&&{}
\eeqra
as should be clear from (\ref{bc}) and (\ref{bcops}).

The factorized equations (\ref{bcfactor}) suggest the following map between
their solutions. Indeed, given that $L_b b(x) = 0$, then 
clearly
\beq\label{btoc}
c(x) = {1\over \om +\pix} Q\, b(x)
\eeq
is a solution of $L_c c(x) = 0$. Similarly, if $L_c c(x) =0$, then 
\beq\label{ctob}
b(x) = {1\over \om -\pix} Q^\dgg \,c(x)
\eeq
solves $L_b b(x) =0$.

Thus, in particular, given a pair $\{b_1(x), b_2(x)\}$ of independent 
fundamental solutions of $L_b b(x) =0$, we can obtain from it 
a pair $\{c_1(x), c_2(x)\}$ of independent 
fundamental solutions of $L_c c(x) =0$ by using (\ref{btoc}), and vice
versa. Therefore, with no loss of generality, we henceforth assume,
that the two pairs of independent fundamental solutions $\{b_1(x), b_2(x)\}$
and $\{c_1(x), c_2(x)\}$, are related by (\ref{btoc}) and (\ref{ctob}).

The coefficients 
$A_{b,c}^{(1)}(k), A_{b,c}^{(2)}(k)$ in (\ref{planewaves}) are to be adjusted
according to (\ref{btoc}) and (\ref{ctob}), and this was the purpose of 
introducing them in the first place. 

Thus, with no loss of generality, we may make the standard choice 
\beq\label{standardchoice}
A_{b}^{(1)} = A_{b}^{(2)} =1
\eeq 
in (\ref{planewaves}). The coefficients $A_{c}^{(1)}, A_{c}^{(2)}$ are 
then determined by (\ref{btoc}):
\beqra\label{Accoeffs}
A_c^{(1)} &=& {\si(-\infty)-ik\over \pi (-\infty) + \om}\nonumber\\
{}\nonumber\\
A_c^{(2)} &=& {\si(\infty)+ik\over \pi (\infty) + \om}\,. 
\eeqra

We note that these $b(x)\leftrightarrow c(x)$ mappings can break only if 
\beq\label{mapbreak}
Q\,b = 0 \quad\quad {\rm or}\quad\quad Q^\dgg\,c =0\,,
\eeq
for $b(x)$ or $c(x)$ that {\em solve} (\ref{bcfactor}).  Do such solutions 
exist? Let us assume, for example, that $Q\,b = 0$ and that $L_b b=0$. From 
the first equation in (\ref{bcfactor}) (or in (\ref{bc})), we see that 
this is possible 
if and only if $\om\pm\pix\equiv 0$, which clearly cannot hold if 
$\pax\pix\neq 0$. A similar argument holds for $Q^\dgg\,c =0$. 
Thus, if $\pax\pix\neq 0$, the mappings (\ref{btoc}) and 
(\ref{ctob}) are one-to-one. In particular, a bound state in $L_b$ 
implies a bound state in $L_c$ (at the same energy) and vice-versa.

An interesting related result concerns the wronskians $W_b$ and $W_c$. 
From (\ref{wronskian}), and from (\ref{btoc}) and (\ref{ctob}) it follows 
immediately that for pairs of independent fundamental solutions 
$\{b_1(x), b_2(x)\}$ and $\{c_1(x), c_2(x)\}$ we have 
\beq\label{wbwc}
W_c = {c_2\pax c_1 - c_1\pax c_2\over \om-\pix} = c_1b_2 -c_2b_1 = 
{b_2\pax b_1 - b_1\pax b_2\over \om+\pix} = W_b\,.
\eeq
The wronskians of pairs of independent fundamental solutions of $L_b$ and 
$L_c$, which are related via (\ref{btoc}) and (\ref{ctob}) are equal!

To summarize, if $\pax\pix\neq 0$, $L_b$ and $L_c$ have the same set of 
energy eigenvalues and their eigenfunctions are in one-to-one correspondence. 

If, however, $\pi=$const., then we are back to the familiar ``supersymmetric'' 
factorization
\beq\label{susyfactor}
Q^\dgg\,Q\, b = (\om^2-\pi^2)\,b\,, \quad\quad 
Q\, Q^\dgg\, c = (\om^2-\pi^2)\,c\,,
\eeq
and mappings 
\beq\label{susymap}
c(x) = {1\over \om +\pi} Q\, b(x)\,,\quad\quad 
b(x) = {1\over \om -\pi} Q^\dgg \,c(x)\,.
\eeq
As is well known from the literature on supersymmetric quantum mechanics, 
the mappings (\ref{susymap}) break down if either $Qb=0$ or 
$Q^\dgg c=0$, in which case the two operators $Q^\dgg Q$ and $QQ^\dgg$ are 
isospectral, but only up to a ``zero-mode'' (or rather, an 
$\om^2=\pi^2$ mode), which belongs to the spectrum of only one of the 
operators\footnote{This is true for short range 
decaying potentials on the whole real line. Strictly speaking, 
(to the best of our knowledge) only the case 
$\pi=0$ appears in the literature on supersymmetric quantum mechanics.}. 
The case $\pix\equiv 0$ brings us back to the GN model. Supersymmetric 
quantum mechanical considerations were quite useful in the study of
fermion bags in \cite{josh1}. 

The ``Witten index'' associated with the pair of isospectral operators 
$L_b$ and $L_c$, is always null for backgrounds 
in which $\pax\pix\neq 0$, since they are absolutely isospectral, 
and not only up to zero modes. There is no interesting topology associated
with spectral mismatches of $L_b$ and $L_c$. This is not surprising at all, 
since the NJL model, with 
its continuous axial symmetry, does not support topological solitons. This is
in contrast to the GN model, for which $\pi\equiv 0$, which contains 
topological kinks, whose topological charge is essentially the Witten index
of the pair of operators (\ref{susyfactor}).

We note in passing that isospectrality of $L_b$ and $L_c$ which we have 
just proved, is consistent with the $\gam_5$ symmetry of the system 
of equations in (\ref{greens}), which relates the resolvent of $D$ with that  
of $\tilde D = -\gam_5 D \gam_5 $. Due to this symmetry, we can map 
the pair of equations $L_b b(x,y) = \delta (x-y)$ and 
$L_c c(x,y) = -\delta (x-y)$ (Eqs. (\ref{bc})) on each other by 
\beq\label{bcflip}
b(x,y)\leftrightarrow -c(x,y)\quad {\rm together~ with}\quad  
(\si,\pi)\rightarrow (-\si,-\pi)\,.
\eeq 
(Note that under these 
reflections we also have $a(x,y)\leftrightarrow -d(x,y)$, as we can see 
from (\ref{ad}).) The reflection $(\si,\pi)\rightarrow (-\si,-\pi)$ just 
shifts both asymptotic chiral angles $\theta_\pm$ by the same amount 
$\pi$, and clearly does not change the physics. Since this reflection 
interchanges $b(x,y)$ and $c(x,y)$ without affecting the physics, these two 
objects must have the same singularities as functions of $\om$, consistent 
with isospectrality of $L_b$ and $L_c$.

\subsection{The Diagonal Resolvent}

Following \cite{josh2,fz} we define the diagonal resolvent 
$\langle x\,|iD^{-1} | x\,\rangle$ symmetrically as
\beqra
\langle x\,|-iD^{-1} | x\,\rangle &\equiv& \left(\begin{array}{cc} A(x) & 
B(x) \\{}&{}\\ C(x) & 
D(x)\end{array}\right)\nonumber\\{}\nonumber\\{}\nonumber\\
 &=& {1\over 2} \lim_{\epsilon\rightarrow 0+}\left(\begin{array}{cc} 
a(x,y) + a(y,x) &  b(x,y) + b(y,x)\\{}&{}\\ c(x,y) +c(y,x) & d(x,y) + 
d(y,x)\end{array}\right)_{y=x+\epsilon}
\label{diagonal}
\eeqra
Here $A(x)$ through $D(x)$ stand for the entries of the diagonal 
resolvent, which following (\ref{ad}) and (\ref{bcexpression}) have the 
compact representation\footnote{$A, B, C$ and $D$ are obviously functions 
of $\omega$ as well. For notational simplicity we suppress their explicit 
$\omega$ dependence.}
\beqra
B(x)&=&~~{b_1(x)b_2(x)\over W_b}\quad\quad , \quad\quad D(x)={i\over 
2}{\left[\pax+2\sigx\right]B\left(x\right)\over 
\omega+\pix}\,,\nonumber\\
C(x)&=&-{c_1(x)c_2(x)\over W_c}\quad\quad , \quad\quad A(x)={i\over 
2}{\left[\pax-2\sigx\right]C\left(x\right)\over \omega-\pix}\,.
\label{abcd}
\eeqra

We now use the generalized ``supersymmetry'' of the Dirac operator,
which we discussed in the previous subsection, to deduce some important 
properties of the functions $A(x)$ through $D(x)$.

From (\ref{abcd}) and from (\ref{qqdagger}) we we have 
$$ A(x) = {i\over 2}{\pax-2\sigx\over \omega-\pix}\left(-{c_1c_2\over W_c}
\right)={i\over 2W_c}{c_2 Q^\dgg c_1 + c_1 Q^\dgg c_2 \over \omega-\pix}\,.$$
Using (\ref{ctob}) first, and then (\ref{btoc}), we rewrite this expression as 
$$A(x) = {i\over 2W_c}(c_2b_1 + c_1b_2) = {i\over 2W_c}  
{b_1 Q b_2  + b_2 Q b_1 \over \omega+\pix}\,.$$
Then, using the fact that $W_c=W_b$ (Eq. (\ref{wbwc})) and (\ref{abcd}),
we rewrite the last expression as 
$$A(x) = {i\over 2}{\pax + 2\si \over \omega+\pix}\left({b_1b_2\over W_b}
\right) = {i\over 2}{(\pax + 2\si)B(x) \over \omega+\pix}\,.$$
Thus, finally, 
\beq\label{ADeq}
A(x)=D(x)\,.
\eeq
Supersymmetry renders the diagonal elements $A$ and $D$ equal.

Due to (\ref{abcd}), $A=D$ is also a first order differential equation
relating $B$ and $C$. We can also relate the off diagonal elements 
$B$ and $C$ to each other more directly.
From (\ref{abcd}) and from (\ref{btoc}) we find
\beqra\label{cfromb}
C(x) = -{c_1c_2\over W_c} = -{(Qb_1)(Qb_2)\over (\om+\pi )^2 W_c}\,.
\eeqra
After some algebra, and using (\ref{wbwc}), we can rewrite this as 
$$ -(\om+\pi )^2 C = \si^2 B + \si B' + {b_1' b_2' \over W_b}$$
The combination $b_1'b_2'/W_b$ appears in $B''=(b_1b_2/W_b)''$. 
After using $L_b b_{1,2}=0$ to eliminate $b_1''$ and $b_2''$ from $B''$,
we find
$${b_1'b_2'\over W_b} = {1\over 2}B'' - {\pi'B'\over 2(\om+\pi)} - 
\left(\si^2 +\pi^2 - \si' -\om^2 + {\si\pi'\over \om + \pi}\right)B$$
Thus, finally, we have 
\beq\label{BtoCrel}
-(\om+\pi )^2 C = {1\over 2}B'' + \left( \si - {\pi'
\over 2(\om+\pi)}\right)B' - \left(\pi^2 - \si' -\om^2 + 
{\si\pi'\over \om + \pi}\right) B\,.
\eeq
In a similar manner we can prove that 
\beq\label{CtoBrel}
(\om-\pi )^2 B = -{1\over 2}C'' + \left( \si - {\pi'
\over 2(\om-\pi)}\right)C' + \left(\pi^2 + \si' -\om^2 + 
{\si\pi'\over \om - \pi}\right) C\,.
\eeq

We can simplify (\ref{BtoCrel}) and (\ref{CtoBrel}) further. 
After some algebra, and using (\ref{abcd}) we arrive at 
\beqra\label{BCfinal}
C(x) &=& {i\over \om+\pix}\,\pax D(x) - {\om - \pix \over \om + \pix}
\,B(x)\nonumber\\{}\nonumber\\
B(x) &=& {i\over \om - \pix}\,\pax A(x) - {\om + \pix \over \om - \pix}
\,C(x)\,.
\eeqra
Supersymmetry, namely, isospectrality of $L_b$ and $L_c$, enables us 
to relate the diagonal resolvents of these operators, $B$ and $C$, to each 
other. 

Thus, we can use (\ref{abcd}), (\ref{ADeq}) and (\ref{BCfinal})
to eliminate three of the entries of the diagonal resolvent in (\ref{abcd}),
in terms of the fourth. 

Note that the two relations in (\ref{BCfinal}) transform 
into each other under 
\beq\label{diagBCflip}
B\leftrightarrow -C\quad {\rm simultaneously~ with}\quad 
(\si,\pi)\rightarrow (-\si,-\pi)\,,
\eeq
in consistency with (\ref{bcflip}). 
The relations in (\ref{BCfinal}) are linear and homogeneous, with 
coefficients that for $\pax\pix\neq 0$ 
do not introduce additional singularities in the 
$\om$ plane. Thus, we see, once more, that $B$ and $C$ have the same 
singularities in the $\omega$ plane. We refer the reader to 
Section 4 in \cite{fz} for concrete examples of such resolvents.

The case $\pix\equiv 0$ brings us back to the GN model. In the GN model, our 
$B$ and $C$, coincide, respectively, with $\om R_-$ and $-\om R_+$, defined 
in Eqs. (9) and (10) in \cite{josh1}. With these identifications, the 
relation $A=D$ (Eq. (\ref{ADeq})) coincides essentially with Eq. (18) of 
\cite{josh1}. The relations (\ref{BtoCrel}) and (\ref{CtoBrel}) were not 
discussed in \cite{josh1}, but one can verify them, for example, for the 
resolvents corresponding to the kink case $\sigx = m\,{\rm tanh}\, mx$ 
(Eq. (29) in \cite{josh1}), for which $$C=-{\om\over 2\sqrt{m^2-\om^2}}\,,\quad
B=\left[\left({m\, {\rm sech}\, mx\over \om}\right)^2 - 1\right]C\,.$$

\newpage
\section{Bilinear Fermion Condensates and Vanishing of the Spatial 
Fermion Current}
\setcounter{equation}{0}

Following basic principles of quantum field theory, we may write the most 
generic flavor-singlet bilinear fermion condensate in our static background
as
\beqra\label{condensate}
&&\langle \bar\psi_{a\alpha}(t,x)\,\Gamma_{\alpha\beta}
\,\psi_{a\beta}(t,x)\rangle_{{\rm reg}}
=N\int {d\om\over 2\pi }\,\rmtr\left[\Gamma 
\langle x| {-i\over \om\gam^0 + i\gam^1\pax 
-\left(\si + i\pi\gam_5\right)} | x \rangle_{{\rm reg}}\right]
\nonumber\\{}\nonumber\\ 
&&= N\int {d\om\over 2\pi }\,
\rmtr\left\{\Gamma\left[ 
\left(\begin{array}{cc} A(x) & 
B(x) \\{}&{}\\ C(x) & 
D(x)\end{array}\right)\,-\, \left(\begin{array}{cc} A & 
B \\{}&{}\\ C & 
D\end{array}\right)_{_{VAC}}\right]\right\}\,,
\eeqra
where we have used (\ref{diagonal}).
Here $a=1,\cdots ,N$ is a flavor index, and the trace is taken over 
Dirac indices $\alpha, \beta$. As usual, we regularized this condensate by 
subtracting from it a short distance divergent piece embodied here by the
diagonal resolvent
\beq\label{vac}
\langle x\,|-iD^{-1} | x\,\rangle_{_{VAC}} = 
\left(\begin{array}{cc} A & 
B \\{}&{}\\ C & 
D\end{array}\right)_{_{VAC}} = 
{1\over 2 \sqrt{m^2-\omega^2}}
\left(\begin{array}{cc} i m{\rm cos}\theta & \omega +m{\rm sin}\theta 
\\{}&{}\\ -\omega+m{\rm sin}\theta &  i m{\rm 
cos}\theta\end{array}\right)
\eeq
of the Dirac operator in a vacuum configuration $\si_{_{VAC}} = 
m {\rm cos}\theta$ and $\pi_{_{VAC}} = m{\rm sin}\theta$.

In our convention for $\gam$ matrices (\ref{majorana}) we have 
\beq\label{diagresolvent}
\left(\begin{array}{cc} A(x) & 
B(x) \\{}&{}\\ C(x) & 
D(x)\end{array}\right)\, = {A(x)+D(x)\over 2}{\bf 1} + 
{A(x)-D(x)\over 2i}\gam^1 + i{B(x)-C(x)\over 2}\gam^0 + 
{B(x)+C(x)\over 2}\gam_5\,.
\eeq

An important condensate is the expectation value of the fermion current 
$\langle j^\mu (x)\rangle $. In particular, consider its spatial component.
In our static background $(\sigx ,\pix)$, it 
must, of course, vanish identically
\beq\label{j1vanish}
\langle j^1 (x)\rangle =0\,.
\eeq
Thus, substituting $\Gamma = \gam^1$ in (\ref{condensate}) and using 
(\ref{diagresolvent}) we find 
\beq\label{fourierspat}\langle j^1 (x)\rangle = 
iN\int {d\om\over 2\pi }\,\left[A(x)-D(x)\right]\,.
\eeq
But we have already proved that $A(x)=D(x)$ in {\em any} 
static background $(\sigx ,\pix)$ (Eq.(\ref{ADeq})). Thus, each
frequency component of $\langle j^1 \rangle$ vanishes separately, and 
(\ref{j1vanish}) holds identically. 
It is remarkable that the generalized supersymmetry of the Dirac operator
guarantees the consistency of any static $(\sigx ,\pix)$ 
background.

Expressions for other bilinear condensates may be derived in a similar 
manner (here we write the unsubtracted quantities). 
Thus, substituting $\Gamma = \gam^0$ in (\ref{condensate}) and using 
(\ref{diagresolvent}), (\ref{ADeq}) and (\ref{BCfinal}), 
we find that the fermion density is 
\beq\label{fourierdensity}
\langle j^0 (x)\rangle = 
iN\int {d\om\over 2\pi }\,\left[B(x)-C(x)\right] = 
iN\int {d\om\over 2\pi }\,{2\om B(x) -i\pax D(x)\over  \om + \pix}\,.
\eeq
Similarly, the scalar and pseudoscalar condensates are 
\beq\label{fourierscalar}
\langle \bar\psi (x)\psi (x)\rangle = 
N\int {d\om\over 2\pi }\,\left[A(x)+D(x)\right] = 
2N\int {d\om\over 2\pi }\, D(x)\quad\quad \,,
\eeq
and 
\beq\label{fourierpseudoscalar}
\langle \bar\psi (x)\gam^5\psi (x)\rangle = 
N\int {d\om\over 2\pi }\,\left[B(x)+C(x)\right] =
N\int {d\om\over 2\pi }\,{2\pix B(x) +i\pax D(x)\over  \om + \pix}\,.
\eeq

\vspace{2cm}
{\bf Acknowledgements}~~~ I am happy to thank A. Zee and 
M. Moshe for useful discussions. This work has been supported in 
part by the Israeli Science Foundation grant number 307/98 (090-903).


\newpage
\begin{thebibliography}{99}

\bibitem{sphericalbag} T. D. Lee and G. Wick, Phys. Rev. D {\bf 9}, 2291 
(1974);\\
R. Friedberg, T.D. Lee and R. Sirlin, Phys. Rev. D {\bf 13}, 2739 
(1976);\\
R. Friedberg and T.D. Lee, Phys. Rev. D {\bf 15}, 1694 (1976), {\em 
ibid.} 
{\bf 16}, 1096 (1977);\\
A. Chodos, R. Jaffe, K. Johnson, C. Thorn, and V. Weisskopf,  Phys. Rev. 
D {\bf 9}, 3471 (1974).

\bibitem{shellbag} W. A. Bardeen, M. S. Chanowitz, S. D. Drell, M. 
Weinstein 
and T. M. Yan, Phys. Rev. D {\bf 11}, 1094 (1974);\\
M. Creutz, Phys. Rev. D {\bf 10}, 1749 (1974).

\bibitem{mackenzie} R. MacKenzie, F. Wilczek and A.Zee,  Phys. Rev. Lett 
{\bf 53}, 2203 (1984).

\bibitem{gn}  D.J. Gross and A. Neveu,  Phys. Rev. D  {\bf 10},   3235
(1974).

\bibitem{njl} Y. Nambu and G. Jona-Lasinio, Phys. Rev. {\bf 122}, 345 
(1961), {\it ibid} {\bf 124}, 246 (1961).

\bibitem{ccgz} C.G. Callan, S. Coleman, D.J. Gross and A. Zee,
unpublished; This work is described by 
D.J. Gross in {\sl Methods in Field Theory\/}, R. Balian and J. 
Zinn-Justin (Eds.), Les-Houches session  XXVIII 1975 (North Holland, 
Amsterdam, 1976).

\bibitem{dhn}  R.F. Dashen, B. Hasslacher and A. Neveu,  Phys. Rev. D 
{\bf 12}, 2443 (1975).

\bibitem{shei} S. Shei,  Phys. Rev. D {\bf 14}, 535 (1976).

\bibitem{others} See e.g., A. Klein, Phys. Rev. D {\bf 14}, 558 (1976);\\
R. Pausch, M. Thies and V. L. Dolman, Z. Phys. A {\bf 338}, 441 (1991).

\bibitem{josh1} J. Feinberg,  Phys. Rev. D {\bf 51}, 4503 (1995).

\bibitem{fz}  J. Feinberg and A. Zee,  Phys. Rev. D 
{\bf 56},5050 (1997); Int. J. Mod. Phys. {\bf A12}, 1133 (1997). 

\bibitem{massivegn} J. Feinberg and A. Zee,  Phys. Lett. 
{\bf B411}, 134 (1997)

\bibitem{thies} V. Schoen and M. Thies, {\sl 2d Model Field Theories at 
Finite Temperature and Density}, Contribution to the Festschrift in honor of 
Boris Ioffe, (M. Shifman Ed.), hep-th/0008175.

\bibitem{jaffe} E. Farhi, N. Graham, R.L. Jaffe and H. Weigel, Nucl. Phys. 
{\bf B585}, 443 (2000);  Phys. Lett. {\bf B475}, 335 (2000).

\bibitem{bashinsky} S. V. Bashinsky, Phys. Rev. D {\bf 61}, 105003 (2000).
 
\bibitem{stone} I. Kosztin, S. Kos, M. Stone and A. J. Leggett, 
Phys. Rev. B {\bf 58}, 9365 (1998); S. Kos and M. Stone, Phys. Rev. B 
{\bf 59}, 9545 (1999).

\bibitem{faddeev}  L.D. Faddeev and L.A. Takhtajan, {\sl Hamiltonian 
Methods
in the Theory of Solitons\/} (Springer Verlag, Berlin, 1987).\\
L.D. Faddeev, J. Sov. Math. {\bf 5} (1976) 334. This paper is reprinted 
in\newline
L.D. Faddeev, {\sl 40 Years in Mathematical Physics} (World Scientific, 
Singapore, 1995). \\   
S. Novikov, S.V. Manakov, L.P. Pitaevsky and V.E. Zakharov, {\sl Theory of 
Solitons - The Inverse Scattering Method\/} 
(Consultants Bureau, New York, 1984)\newline
( Contemporary Soviet Mathematics). 



\bibitem{josh2} J. Feinberg, Nucl. Phys. {\bf B433}, 625 (1995).

\end{thebibliography}
\end{document}